\documentclass[prl,aps,showpacs,epsf,twocolumn]{revtex4}

\usepackage{graphicx}
\usepackage{epsfig}
\usepackage{amsmath}
\usepackage{amssymb}
\usepackage{bm}

\begin{document}
\title{Total control over ultracold interactions via electric and magnetic fields}
\author{Bout Marcelis, Boudewijn Verhaar, and Servaas Kokkelmans}
\affiliation{Eindhoven University of Technology, P.O.~Box~513, 5600~MB  Eindhoven, The Netherlands}
\date{October 2, 2007}

\hyphenation{Fesh-bach}

\begin{abstract}
The scattering length is commonly used to characterize the strength of ultracold atomic interactions, since it is the leading parameter in the low-energy expansion of the scattering phase shift. Its value can be modified via a magnetic field, by using a Feshbach resonance. However, the effective range term, which is the second parameter in the phase shift expansion, determines the width of the resonance and gives rise to important properties of ultracold gases. Independent control over this parameter is not possible by using a magnetic field only. We demonstrate that a combination of magnetic and electric fields can be used to get independent control over both parameters, which leads to full control over elastic ultracold interactions.
\end{abstract}


\maketitle

Cold atomic gases with resonant interactions are extremely versatile systems. They allow for the experimental realization of a variety of fundamental models from condensed matter or high energy physics, and can go well beyond the basics of those models.
Illustrative examples are the experiments on the crossover between BCS- and BEC-type superfluidity with fermionic gases~\cite{regal04,chin04,bourdel04,kinast04,zwierlein05,zwierlein06,partridge06}, and the experimental signature of the long-sought three-body Efimov trimer states with bosonic gases~\cite{kraemer06}. Also, mixtures of fermionic and bosonic atoms lead to interesting possibilities: heteronuclear molecules can be created with anisotropic interactions, for which new quantum phases have been predicted \cite{ospelkaus06,marcelis07,doyle04,dulieu06}.

Key to these exciting developments is the ability to precisely control the interatomic interactions in ultracold alkali-metal gases. Magnetic field induced Feshbach resonances \cite{feshbach58} have been the indispensable tool providing experimental control of the strength of the interactions. The low-energy interactions are characterized by the s-wave scattering length and can be tuned from weak to strong and from repulsive to attractive by simply changing the magnetic field. Magnetic field sweeps over a Feshbach resonance have been used to create ultracold molecules \cite{donley02,herbig03,xu03,durr04}.

The width of the Feshbach resonance is an important parameter that is directly related to the energy dependence of the interactions. In the case of a broad resonance, the scattering length can be used to describe the system at all relevant energy scales, leading to universal behavior \cite{Heiselberg,Carlson,ho,ohara}. However, in the case of a narrow Feshbach resonance the microscopic physics underlying the resonance can give rise to non-trivial energy dependence and additional scattering parameters are needed to describe the system \cite{combescot,palo,bruun,marcelis04}. The energy dependence of the interaction strength may lead to interesting new physics such as the localisation of Efimov states \cite{petrov}, phonon-like induced superfluidity in fermionic gases \cite{marcelis06}, as well as a clearly distinct Higgs mode in the excitation spectrum of superfluid fermions \cite{barankov}.

A second way of tuning the low energy properties of these systems is to apply a dc electric field \cite{marinescu98,deb01,krems06,li07}. The electric field polarizes the colliding atoms and induces an electric dipole-dipole interaction. This anistropic interaction couples states of different orbital angular momentum. The coupling can give rise to potential or Feshbach resonances in the s-wave scattering length at particular electric field values.

In the presence of either an electric or a magnetic field, the zero-energy scattering length is the only parameter that can be independently tuned. However, we demonstate that for a suitable combination of electric and magnetic fields the width of the Feshbach resonance can also be controlled. In this work we focus on the interplay between the electric field induced potential resonances and magnetic field induced Feshbach resonances, which leads to the possibility of tailoring the energy dependence of the interactions.

This work is organised as follows: We start by explaining the physics underlying the interesting interplay between the potential and Feshbach resonances, as induced and controlled by electric and magnetic fields. We indicate how a simple analytical model captures the essential physics and compare the model to full coupled-channels calculations. Since rubidium is currently used in electric trap experiments~\cite{schlunk,rieger}, we give several examples of the interaction properties of $^{85}$Rb atoms. It can be considered as a model system used to give a proof-of-principle of the tunability of the energy dependence of the interactions. We end with concluding remarks.

Feshbach resonances originate from the coupling of two atoms interacting via an energetically open channel, to one (or several) bound state(s) in energetically closed channels. The position of the bound state(s) can be tuned by changing the magnetic field. When such a bound state is energetically close to the relative kinetic energy of the two atoms, the scattering process becomes resonant. The signature of a potential resonance is a large (absolute) value of the background scattering length, much larger than the range of the van der Waals potential $r_{\rm VDW}=(m C_6/16 \hbar^2)^{1/4}$ \cite{gribakin}, with $m$ the atomic mass and $C_6$ the van der Waals dispersion coefficient. In previous work \cite{marcelis04} we derived a conceptually simple model that encapsulates all the essential physics of the open and closed channel resonance states, and gives a very accurate description of, for instance, the molecular energies and scattering phase shifts for all energies and magnetic fields of interest. At zero energy, the scattering properties are conveniently summarized by the s-wave scattering length
\begin{equation}
a(B)=r_0+a^P\left(1-\frac{\Delta B}{B-B_0}\right),
\end{equation}
where $r_0\simeq r_{\rm VDW}$ is the non-resonant and $a^P$ the resonant part of the open-channel (or background) scattering length. The Feshbach resonance is located at the field $B_0$ and is furthermore characterized by the width $\Delta B$. There is a fundamental relation between the width of the resonance $\Delta B$ and the energy dependence of the interatomic interactions, described by the scattering phase shift \cite{marcelis06}:
\begin{equation}\label{phaseshift}
\delta(k) = -k r_0-\arg{\left( \zeta_k \right)}-\arg{\left(E-\nu + iCk\zeta^{-1}_k \right)},
\end{equation}
where $E=\hbar^2k^2/m$ is the relative collision energy of the atoms in the open channel, $\zeta_k = 1+ika^P$, $\nu = \Delta \mu^{\rm mag}(B-B_0)$ the detuning of the closed-channel bound state and $C=a^P \Delta \mu^{\rm mag} \Delta B$ proportional to the width of the resonance, with $\Delta \mu^{\rm mag}$ the magnetic moment difference between the open-channel scattering threshold and the closed-channel bound state. At the resonant field $B_0$, the resonance width can be related to a length scale that plays the role of an effective range parameter \cite{marcelis06}:
\begin{equation}
R^{\ast}(B) = \frac{\hbar^2}{m (C -a^P\nu)}\zeta_k.
\end{equation}
In the case that $R^{\ast}\ll r_0$ and $\sqrt{R^{\ast}a^P} \ll r_0$, the resonance is called broad and the phaseshift is well described by the simple relation $\delta(k) = -\arctan{[k a(B)]}$ for the energies $k\ll r_0^{-1}$ of interest to cold collisions. If $R^{\ast}\gg r_0$ and/or $\sqrt{R^{\ast}a^P} \gg r_0$, the resonance is narrow and energy dependent corrections beyond the scattering length approximation are needed to properly describe the interactions. In all cases Eq.~(\ref{phaseshift}) gives an accurate description at all relevant energies.

In the presence of an electric field, the electronic s-wave symmetry of the individual atoms is distorted due to the electric dipole coupling with excited electronic p-wave states. At sufficiently weak fields the atoms acquire a dipole moment according to ${\bf d_i} = \alpha_i(0) {\bm \varepsilon}$, where $\alpha_i(0)$ is the static electric polarizability of atom $i$, and $\bm \varepsilon$ is the applied electric field. The interaction between the electric dipoles takes the familiar form \cite{messiah}:
\begin{equation}
\frac{\alpha_1(0) \alpha_2(0)}{4\pi \epsilon_0 r^3} \left[ {\bm \varepsilon \cdot \bm \varepsilon} - 3\left( {\bm \varepsilon \cdot \bf \hat{r} } \right)\left( {\bm \varepsilon \cdot \bf \hat{r} } \right) \right],
\end{equation}
where $\epsilon_0$ is the vacuum dielectric constant. The angle-dependent structure of the anistropic interactions can be rewritten in terms of a scalar product of two irreducible spherical tensors of rank 2. Due to its tensorial nature, the anisotropic part of the interaction obeys triangle type selection rules and couples states with orbital angular momentum according to the selection rule $l-l'=0,2$, with the exception of $l=0 \rightarrow l'=0$ \cite{messiah}.

When an electric field is applied, the coupling between different angular momentum states has several important effects. First of all, the energy of the weakliest bound (or virtual) state in the open channel is shifted, and the sign and strength of the open-channel scattering, characterized by $a^P = a^P(\varepsilon)$, can be tuned in and out of resonance (see inset Fig.~2). This also allows for control of the energy dependence of the interactions, characterized by $R^{\ast}$. It can be shown that the width of the resonance is proportional to the resonant part of the open-channel scattering length, $\Delta B \propto a^P$ and therefore $C \propto (a^P)^2$. The effective range parameter $R^{\ast}$ can be controlled through its dependence on $a^P$ and $C$. Furthermore, the energies of the s-wave bound states in the closed channels are shifted, which leads to a shift in the positions of the Feshbach resonances, $B_0 = B_0(\varepsilon)$, and additional (usually very narrow) resonances show up due to the electric field induced coupling of the open-channel scattering state to closed-channel bound states.

Using the electric field, the interactions at non-zero energies can be controlled. The magnetic field can be used independently to control the total scattering length $a(B,\varepsilon)$ at this particular electric field value. This independent control of the zero-energy interactions, and the non-zero energy corrections, provides the unique opportunity to tailor the interatomic interactions in many possible ways. In practice, this means full control over the elastic interactions in the ultracold regime. We give interesting examples in the remainder of this work.

We use a full coupled-channels method~\cite{cc} for a rigorous determination of the scattering properties of the Rb atoms. The spin-spin interaction in this method comes from the magnetic dipole and second-order spin-orbit interactions, and its angular structure is very similar to the electrically induced dipole-dipole interaction, which has its origin in the Stark effect. We extended the existing numerical code by carefully including the electric dipole-dipole interactions, allowing for arbitrary relative orientations of the applied magnetic and electric fields. Interestingly, to a high degree the two fields operate on independent degrees of freedom: $\bf B$ influences the spin system and $\bm \varepsilon$ the orbital system. Each of these subsystems has its rotational symmetry axis, the directions of $\bf B$ and $\bm \varepsilon$, respectively, and therefore its own conserved quantum numbers ($m_F \parallel \bf B$ and $m_l \parallel \bm \varepsilon$). Only the relatively weak spin-spin interaction correlates the two parts. Varying the orientation of $\bf B$ or $\bm \varepsilon$ in future experiments might therefore offer fascinating prospects for studying quantum degenerate gas samples with different preferred electric and magnetic axes. From this point of view $^{87}$Rb might be a promising choice, since its d-wave shape resonance may be expected to enhance collective quadrupole oscillations of atom pairs induced by time-varying magnetic or electric fields.

We now consider $^{85}$Rb atoms in the $(f_1,f_2)F,m_F = (2,2)4,-4$ two-particle hyperfine state, where $f_i$ are the hyperfine quantum numbers of the individual atoms that add up to total spin $F$ with projection $m_F$ on the magnetic axis. The electric field is taken parallel to the magnetic field. In this case, there is only one s-wave channel that is energetically open. We calculated the scattering length for this channel as a function of the magnetic field at several electric field strengths, and typical results are shown in Fig.~1. In principle, the coupling to $l\neq 0$ open channels can give rise to inelastic losses. However, in the cases we consider here the effect is very small and can be neglected.

\begin{figure}
\includegraphics[width=0.8\columnwidth]{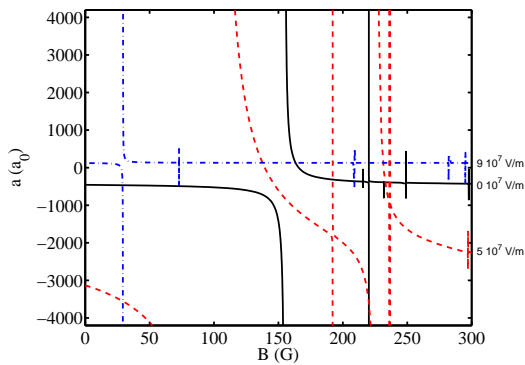}
\caption{(Color online) $s$-wave scattering length $a$ as a function of the magnetic field at three different electric field strengths.}
\end{figure}

In the absence of an electric field, there are several Feshbach resonances within the range of magnetic fields shown, as seen from the continuous line of Fig.~1. The broad resonance at $B_0 = 154.85$ G is related to an $s$-wave molecular state with total spin quantum number $F=4$, while there is a narrow resonance at $B_0 \simeq 220.20$ G that originates from an $F=6$ $s$-wave molecular state. The other resonances shown are extremely narrow and originate in the anisotropic coupling to $F=6$ closed-channel bound states with $l=2$. The large and negative background scattering length $a_{\rm bg} =r_0+a^P$ consists of a non-resonant part $r_0\simeq 120$ a$_0$ and a resonant part $a^P \simeq -570$ a$_0$ related to a virtual state. Examples of the scattering length at non-zero electric fields are shown as the dashed and dash-dotted line. It is clear that the electric field has a large effect on the resonance structure. The Feshbach resonances that were considered initially at zero field are shifted towards lower or higher magnetic fields and their widths are increased or decreased by orders of magnitude.

\begin{figure}
\includegraphics[width=0.75\columnwidth]{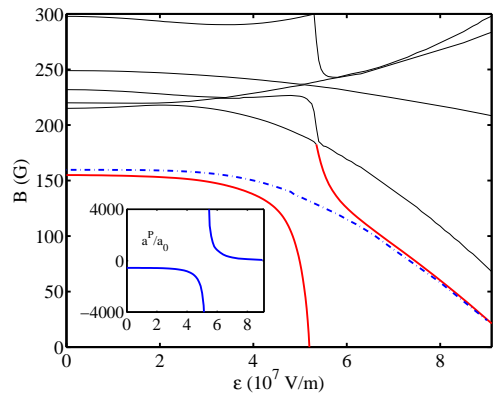}
\caption{(Color online) Position $B_0$ of several resonances as a function of magnetic and electric fields. See text for details. Inset: resonant part $a^P$ of open channel scattering length as a function of electric field.}
\end{figure}

In Fig.~2 we show the magnetic field position of the broadest resonance (thick red line) as a function of the applied electric field. The position of the bare closed-channel state, shifted from the (dressed) resonance position by the width as $B'_0=B_0-\Delta B$ \cite{marcelis04}, is shown as the dash-dotted line. The narrow resonances are indicated by thin lines. If the electric field is increased, the bare states are shifted due to their anisotropic coupling with other closed-channel states. Simultaneously, the virtual state in the open channel, coupled to $l\neq 0$ open-channel states, approaches the scattering threshold and becomes bound at the resonance field of about $\varepsilon_r \simeq 5.3 \cdot 10^7$ V/m. This gives rise to a divergence in the open channel scattering length $a^P$, as clearly visible in the inset of Fig.~2. For $\varepsilon > \varepsilon_r$, $a^P$ changes sign and for high electric fields the open channel eventually becomes non-resonant.

From Fig.~2 it is evident that the electric field induced couplings give rise to an avoided crossing between the open-channel shape resonance and the Feshbach resonance. We checked that this avoided crossing is accurately described by the analytical scattering model of Ref.~\cite{marcelis04}, which predicts the scaling $\Delta B \propto a^P$. Using this model, the numerical data of Fig.~2 (thick red line) are accurately reproduced.

From these considerations, we reach the exciting conclusion that the width of the resonance can be simply controlled by varying the electric field. It can be increased when approaching the resonance in $a^P$ at $\varepsilon_r$. When $a^P$ becomes non-resonant at high electric fields, the Feshbach resonance becomes narrower. An example is shown by the dash-dotted line of Fig.~1, where the scattering length shows a narrow resonance at a field of $B_0 = 26.1$ G, which is related to the same $F=4$ molecular state that gives rise to the broad $154.9$ G resonance at zero electric field.

Perhaps the most interesting situation to study is a narrow resonance with a resonant open channel. In the vicinity of $\varepsilon_r$, where $a^P$ can be tuned to large positive or negative values, we therefore focus our attention on one of the additional narrow resonances that are coupled in by the electric field. We fix the electric field just above the electrically induced resonance at $\varepsilon = 5.6\cdot 10^7$ V/m. This gives a large and positive $a^P = 1916$ a$_0$ (see inset Fig.~2), related to a weakly bound state in the open channel. The magnetic field is tuned close to the narrow resonance at $B_0 = 179.50$ G. At the left side of the resonance, where there are no molecules in the system, the total scattering length is first large and negative, but has a zero crossing at $179.48$ G. Just to the left of this zero crossing, the scattering length becomes positive and the zero-energy interactions have a repulsive character. However, due to the fact that this resonance is very narrow, the interactions become attractive at non-zero energies, which can for instance lead to the formation of a BCS-like state with positive scattering length \cite{marcelis06}. The tunability of the energy dependence of the interactions is shown in Fig.~3.

\begin{figure}
\includegraphics[width=0.7\columnwidth]{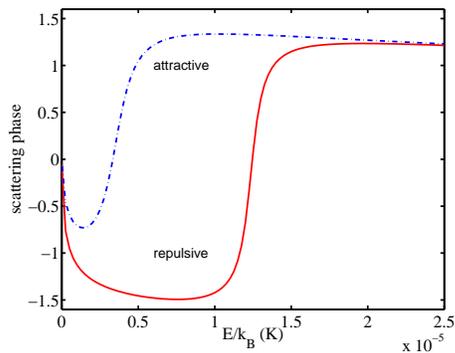}
\caption{(Color online) Example of the tunability of the scattering phase shift by a suitable combination of magnetic and electric fields. Solid line: $B=179.4$ G, dash-dotted line: $B=179.46$ G.}
\end{figure}

In conclusion, we showed that the energy dependence of the interactions can be controlled by combining electric and magnetic fields. This is particularly interesting in two-component fermionic systems, where the electric field could potentially be used to control the crossover from universal behavior to non-universal behavior by tuning the Feshbach resonance from broad to narrow. Tunability of the open-channel scattering length $a^P$ allows for a systematic study of the interplay between open-channel shape resonances and closed-channel Feshbach resonances. Another exciting application is in the study of three-body Efimov physics, where the control of the effective range parameter $R^{\ast}$ gives rise to a tunable three-body parameter \cite{petrov}.

In this work, we used $^{85}$Rb to give a proof-of-principle of the tunability of the relevant interaction parameters, such as the width and open-channel (background) scattering length. A disadvantage of this particular system is the relatively high electric field of order 500 kV/cm needed to achieve the full control over the ultracold interactions. However, the effect is very general and similar control can be achieved in ultracold gases of other alkali-metal atoms. Especially a mixture of different species, in particular light and heavy atoms. Good candidates are $^6$Li fermions with $^{40}$K fermions or $^{133}$Cs bosons, as the permanent dipole moment of the heteronuclear collision complexes gives rise to much lower necessary electric fields of order of tens of kV/cm, easily generated in current experiments \cite{krems06,li07}. The necessary fields of order of hundreds of kV/cm could perhaps be reached in experiments utilizing atom chips.

We thank P.~Pinkse for usefull discussions. This work was supported by the Netherlands Organization for Scientific Research (NWO).

\end{document}